\documentclass[a4paper]{article}
\usepackage{graphicx,amsmath}
\usepackage[T1]{fontenc}
\usepackage[latin2]{inputenc}

\usepackage{latexsym,amsfonts,amssymb,amsmath,amsthm}
\usepackage{fancyhdr}
\usepackage{lmodern}
\DeclareMathAlphabet{\mathbk}{OT1}{cmr}{bx}{sl}

\usepackage{eufrak}

\newtheorem{all}{Proposition}[section]

\newcommand\nc{\newcommand}
\nc\ee{\mathbk e}
\nc\e{\EuFrak e}
\nc\E{\mathbk E}
\nc\aS{\mathtt S}
\nc\Sd{\mathbf S}
\nc\la{\lambda}
\nc\Ll{\mathbk L} \nc\Pf{\boldsymbol{\mathcal P}}\nc\Mf{\boldsymbol{\mathcal M}}
\newcommand\1{\boldsymbol 1}\nc\0{\boldsymbol 0}
 \newcommand\rr{\mathbb R}
\newcommand\vf{\varphi} \newcommand\T{\mathbk T}

\newcommand\dd{\mathbk d}

\nc\beq{\begin{equation}}\nc\enq{\end{equation}}
\nc\x{\mathbk x} \nc\y{\mathbk y}\nc\z{\mathbk z}
\nc\ts{\mathbk t}
\nc\intl{\int\limits}
\nc\tm{\mathrm{tm}}
\nc\n{\mathbk n}\nc\nb{\mathbk n}
\nc\aI{\mathrm I}
\nc\pt{\partial}
\nc\rs{\mathbk r}
\nc\vv{\mathbk v}\nc\hh{\mathbk h}
\newcommand\aM{\mathrm M}\newcommand\M{\mathbf M}
\nc\I{\mathbb T}
\nc\uu{\mathbk u}\nc\U{\mathbk U}
\nc\B{\mathbk B} \nc\F{\mathbk F}
\nc\ii{\mathbk i}\renewcommand\ij{\EuFrak i}
\nc\ran{\mathrm{Ran}}
\nc\jj{\EuFrak j}\nc\ji{\mathbk j}
\nc\Dd{\mathrm D}
\nc\s{\mathbk s}
\nc\q{\mathbk q}
\nc\ad{\mathbk a}
\nc\Supp{\mathrm{Supp}}
\nc\R{\mathbk R}
\nc\ord{\mathrm{Ordo}(\rho)}
\nc\app{\approx}
\nc\otm{\otimes}
\nc\sk{\smallskip}
\nc\dZ{\mathbk Z}
\nc\A{\mathbk A}
\nc\Ef{\boldsymbol{\mathcal E}}
\nc\Bf{\boldsymbol{\mathcal B}}
\nc\Lf{\boldsymbol{\mathcal L}}
\nc\Ff{\boldsymbol{\mathcal F}}
\nc\Tf{\boldsymbol{\mathcal T}}
\nc\Kf{\mathcal K}
\nc\f{\mathbk f}
\nc\Df{\boldsymbol{\mathcal D}}\nc\Hf{\boldsymbol{\mathcal H}}
\nc\Ht{\mathbk H}\nc\Dt{\mathbk D}
\nc\pp{\mathbk p}\nc\Pt{\mathbk P}
\nc\mqed{\blacksquare}
\nc\Mt{\mathbk M}
\nc\Vf{\mathcal V}
\nc\Nf{\boldsymbol{\mathcal N}}\nc\Nt{\mathbk N}
\nc\Gf{\boldsymbol{\mathcal G}}

\begin{document}

\centerline{\Large{\bf On the Electromagnetic Energies and Forces}}
\medskip

\centerline{T. Matolcsi\footnote{Department of Applied Analysis and Computational Mathematics,
E\"otv\"os Lor\'and University, Budapest, Hungary}}

\begin{abstract}

The problem of the `infinite energy' of a point charge is well known in connection with the Lorentz--Abraham--Dirac 
equation and,  more significantly, in quantum electrodynamics. Though it is not stated usually, this is 
strongly related to the old problem of what the electromagnetic energy-momentum tensor is. 

Electromagnetic energies and forces will be examined thoroughly from a new point of view in 
the framework of Distribution Theory.

\end{abstract}

\section*{Introduction}

The question of what the electromagnetic energy and electromagnetic energy flow  are 
is as old as the theory of electromagnetism. 
First, the original definition of the electromagnetic energy flow, the Poynting vector, seems
to be problematic (see e.g. \cite{Bergman},\cite{Lombardi},\cite{CamJim}). 
Then the Abraham-Minkowksi controversy regarding the energy-momentumm tensor is
an open problem still (see \cite{Leo06a} and the references therein).

Similarly, the question of what the electromagnetic force and the electromagnetic interaction are is an open
problem. There are a number of attempts to reinterpret or replace the LAD equation 
(e.g. \cite{Spohn},\cite{Rohrlich},\cite{Gral-Har-Wald}, \cite{lordir}, \cite{RajRaj08a},\cite{Bild}).
In the paper \cite{Mato}, the radiation reaction force and its role in the LAD equation are clarified by  
showing that not a mathematically incorrect derivation of the radiation reaction force  but a 
physical misapprehension is the source of why the LAD equation does not work well.

The role of electromagnetic energies and forces in connection with
continuous media and thermodynamics is an essential question from a
practical, technical point of view, too (see e.g.
\cite{Mau88b},\cite{Van98a},\cite{Bar10a},\cite{Wan17a}). Therefore, it is important
to see clearly what is right and what is not right in the usual approaches,

{\it The present paper deals with energies and forces in general, it tries to
reveal the roots of the problems, opening so a new perspective which can help to find a solution 
(if it exists at all).}

Since Dirac, some applications of Distribution theory appeared in electromagnetism 
(e.g. \cite{Taylor},\cite{Rowe},\cite{Gral-Har-Wald}). The present paper is based upon  
a systematic and over all use of Distribution theory.

The coordinate-free formulation of spacetime expounded in \cite{Matolcsi} is used which makes 
formulas shorter and more easily comprehensible; a brief summary of the fundamental notions and some 
special notations can be found in \cite{Mato}.

The word `distribution' will appear in two different senses: 1. having its everyday meaning, 2. being a 
mathematical notion. To distinguish between them, I write distribution for the first one and Distribution for 
the second one.

The usual setting of Distribution Theory is based on $\rr^n$ 
(\cite{Gelf-Shil},\cite{Demidov},\cite {Dijk}, \cite{Horvath}).
It is a quite simple generalization that   
spacetime and an observer space are taken instead of $\rr^4$ and $\rr^3$.
Another simple generalization is that vector and tensor Distributions are included, too.
All the electromagnetic quantities are considered to be Distributions, denoted by calligraphic letters, $\Ef$, $\Bf$ etc, 
and the guiding principle is that only formulae definable in Distribution Theory can make sense. Nevertheless, the present 
article can be understood without a thorough knowledge of Distributions; besides the elementary notions the only  
non-trivial one is {\bf pole taming}, described in the Appendix.

\section{Basic notions of electrostatics}

The usual and well-known formulae of electrostatics (\cite{Jackson},\cite{Groot}) are reviewed from the point of 
view of Distributions in such a manner that $c=1$, $\hbar=1$ and the electric charge is measured by real numbers.

Statics: all the quantities are time independent in a {\it uniquely determined} standard inertial frame; 
the space of that frame, denoted by $\aS$, is a three dimensional affine space over the Euclidean  vector space $\Sd$.

A static electric field is supposed to be a vector Distribution $\Ef$ in $\aS$; its regular part is called field function and
is denoted by $\E$; for the sake of simplicity, without danger of confusion, we often say field instead of field function, too.

A static charge distribution is a signed measure $\e$, a static dipole distribution is a vector measure $\Pf$, on $\aS$. 
As  Distributions, they act on test functions $\psi$ by integration, e.g. 
$$(\e\mid \psi):=\intl_{\aS}\psi(q)\ d\e(q).$$

The electric field $\Ef$ produced by a given static charge distribution $\e$ and dipole distribution $\Pf$
satisfies the Maxwell equations 
$$\nabla\cdot\Ef=\e - \nabla\cdot\Pf, \qquad \nabla\times\Ef=0.$$

Special cases are when the charge distribution has a continuous density $\rho$,
the dipole distribution has a continuously differentiable density $\Pt$, 
and the electric field is the regular Distribution corresponding to a continuously differentiable function $\E$; then 
$$\nabla\cdot\E=\rho - \nabla\cdot\Pt, \qquad \nabla\times\E=0.$$
holds as well. 

It follows from Poincaré's lemma in Distribution Theory that there is a potential, a scalar Distribution $\Vf$ such that
$\Ef=-\nabla\Vf$; its regular part, denoted by $V$, is the potential function.

Introducing 
\beq\label{dep}\Df:=\Ef+\Pf, \qquad (\Dt:=\E + \Pt),\enq
the first Maxwell equation can be transformed into the most frequently used form
$$\nabla\cdot\Df=\e, \qquad (\nabla\cdot\Dt=\rho).$$

As concerns a dipole distribution, it can exist independently of the electric field (electret) or can depend
on the electric field in a medium (polarization). Since electrets -- contrary to magnets -- are not familiar in every-day life, 
mostly polarizations are taken in usual treatments, and a relation is given between $\Pt$ and $\E$, e.g. $\Pt=\kappa\E$ 
involving $\Dt=\epsilon\E$. 
By this unfortunate manipulation, one hides the dipole distribution and considers $\Dt$ a proper field quantity though it is not:
it includes material quantities whether an independent or a polarized dipole distribution is in question. It is worth repeating: 
{\it $\Df$ -- and $\Dt$ -- is not a proper field quantity}.
  
The charge distribution of a point charge $e$ in the space 
point $q_o$ is $e\delta_{q_o}$. The produced electric field function is 
\beq\label{ponfu}\E(q):=\frac{e}{4\pi}\frac{q-q_o}{|q-q_o]^3}\enq
which is locally integrable, the electric field $\Ef$ is the corresponding regular Distribution.

The dipole distribution of a point dipole $\pp$ in the space 
point $q_o$ is $\pp\delta_{q_o}$. The produced electric field function is (\cite{Jackson} p.101)
\beq\label{dipfu}\E(q):=\frac1{4\pi}\frac{3(q-q_o)(q-q_o)\cdot\pp}{|q-q_o]^5}-\frac{\pp}{|q-q_o|^3}\enq
which is not locally integrable, the electric field is not a regular Distribution. 
$\E$ has a pole at $q_o$; it can be tamed, and it can be proved that the electric field is
\beq\label{dipfi}\Ef=\tm\E +\frac{\pp}{3}\delta_{q_o}\enq
whose regular part is $\E$.

\section{Extraneous and self}

In some papers the total field is separated into an external field and a 
self-field (e.g. \cite{Bergman},\cite{CamJim}) which seems a useful tool for investigations. The field which is 
external to a charge, however, is a self-field for some other charge.

Therefore we introduce a correspondingly modified definition: two charge distributions are called {\bf extraneous} 
to each other if their supports do not intersect.  Accordingly, an electric field $\Ef$ and a charge distribution 
$\e$ are {\bf extraneous} to each other if $\Ef$ is produced by a charge distribution extraneous to $\e$. 
A similar notion of being extraneous to each other is defined for a charge distribution and a dipole distribution, for 
two dipole distributions, for an electric field and a dipole distribution.

The {\bf self-field} of a charge distribution $\e$ or a dipole distribution $\Pf$ is the electric field produced by 
$\e$ or $\Pf$. 

Now let us consider two charge distributions $\e_1$ and $\e_2$, extraneous to each other. Their self-fields 
$\Ef_1$ and $\Ef_2$ satisfy $\nabla\cdot\Ef_1=\e_1$ and $\nabla\cdot\Ef_2=\e_2$; $\Ef_1$ is extraneous to $\e_2$ 
and $\Ef_2$ is extraneous to 
$\e_1$. Further, $\nabla\cdot(\Ef_1+\Ef_2)=\e_1+\e_2$; thus, $\Ef_1 + \Ef_2$ is the self-field of $\e_1 +\e_2$.

\section{About energies in electrostatics}

\subsection{Extraneous energies}

Analogously to the situation in a Newtonian gravitational field, it is commonly accepted that the work we do
when transporting a point charge $e$ from a space point $q_1$ to the space point $q_2$ on any path in 
the static electric field, {\it extraneous} to the charge, and having the potential function $V$, is the path 
integral of $e\E$ which equals $e\bigl(V(q_2)-V(q_1)\bigr)$ (of course, provided that the electric field is 
defined on the path). 

Such a definition, however, is somewhat incorrect. If the charge moves, then the situation is not static. 
It is an empirical fact that an accelerated charge radiates and reacts on itself, and, contrary to 
the Newtonian gravitational case, the work done is modified by this reaction. Even if the path is a straight 
line, then at least when beginning and ending the motion, the charge must be accelerated, and more 
acceleration occurs necessarily if the path is not a straight line. Nevertheless, we can accept this 
formula as an ideal one because the less the acceleration, the less the radiation.

If $V$ tends to zero at infinity, then  
\beq\label{fst} eV(q_o),\enq
provided that $V$ is defined at $q_o$, is the work done when transporting the point charge $e$  
from the infinity to the point $q_o$. This is considered the electric energy 
of the point charge $e$ in the {\it extraneous} electric potential function $V$. 

From that, it is quite obvious that $\rho V$ can be accepted as the static electric energy density
of a charge density $\rho$ in an {\it extraneous} static electric potential function $V$, provided that
$\rho V$ is locally integrable i.e. it defines a regular Distribution; even, it is expected to be 
integrable in order that the total energy be finite.

It is commonly accepted, too, (\cite{Jackson} p.101) that the static electric energy of a point dipole $\pp$ in the space point 
$q_o$ in the {\it extraneous} electric field $\E$ is $-\pp\cdot\E(q_o)$, provided that $\E$ is defined at $q_o$.
Then  $-\Pt\cdot\E$ can be accepted as the static electric energy density 
of a dipole density $\Pt$ in an {\it extraneous} static electric field function $\E$, provided that $\Pt\cdot\E$ is
locally integrable (and even integrable).

Note the important fact that this notion of energy in an extraneous field is a consequence that only an 
{\bf action} is taken into account instead of an interaction: a field acts on  charges and dipoles.
The field, however, is produced by some other charges and/or dipoles and charges and dipoles 
{\bf interact} through fields.

\subsection{Self-energies, charges}\label{sselfen}

The charges inside a static charge distribution interact with each other which results in 
an electric self-energy of the distribution in question. The usual way to get this self-energy runs as follows. 

Let the point charges $e_1,\dots,e_n$ rest at the different space points 
$q_1,\dots,q_n$. Their self-energy is defined as the work done when 
all the charges are transported from infinity to their places.

First step:  no work is done for the first charge because it is carried in an empty space (see \eqref{fst}); no action, 
no interaction, the electric self-energy of a single point charge is zero.

Second step: the charge $e_2$ is transported in the electric field of $e_1$. The work done is
$e_2\frac{e_1}{4\pi|q_2-q_1|}=\frac{e_1e_2}{4\pi|q_2-q_1|}=\frac1{2}\bigl(e_1V_2(q_1)+ e_2V_1(q_2)\bigr)$. 

And so on, after the last step, the work done i.e. the self-energy is
\beq\label{ponten}\frac1{2}\sum_{k=1}^ne_k(V-V_k)(q_k),\enq
where $V$ is the potential produced by all the charges and $V_k$ is the potential produced by the 
$k$-th charge. It is worth emphasizing that this formula, meaningless for a single charge, is valid only for more than one
charge.

In other words, the charge distribution 
\beq\label{pontelo}\sum_{k=1}^ne_k\delta_{q_k},\enq
for $n\ge2$ has the self-energy distribution
\beq\label{pontend}\frac1{2}\sum_{k=1}^ne_k(V-V_k)\delta_{q_k};\enq
their integral is the total charge and the total self-energy.

Note that $e_k(V-V_k)\delta_{q_k}$ is correct because $V-V_k$ is the sum of the potentials of the charges except the $k$-th one. 
$V_k$ is not defined at $q_k$, thus $V_k\delta_{q_k}$ makes no sense according to the original mathematical definition 
of a measure multiplied by a function. The fact that the electric self-energy of a single 
point charge is zero suggests that we define formally $e_kV_k\delta_{q_k}:=0$. Then $V_k$ can be omitted from \eqref{pontend},
and the formula will be meaningful for $n=1$, too.

Putting together more and more particles with smaller and smaller charges, one goes over -- not in a mathematically 
exact sense -- to a continuous charge distribution, thus, without a serious objection, 
\beq\label{conten}\frac1{2}\rho V\enq
{\it can be accepted as
the  self-energy density of a static charge density $\rho$ in its own potential function} $V$, provided that $\frac1{2}\rho V$ is
locally integrable i.e. it defines a regular Distribution; even, it is expected to be 
integrable in order that the total self-energy be finite.

Now it is worth examining the relation between extraneous energy and self-energy. Let us take two charge densities $\rho_1$ 
and $\rho_2$, extraneous to each other, producing the potential functions $V_1$ and $V_2$, respectively. Then 
the charge distributions together have the self-energy\footnote{We use a simplified notation: when integrating by 
the Lebesgue measure (volume form) of $\aS$, $d\la_{\aS}$ is not written in the integrals}

$$\frac1{2}\intl_\aS(\rho_1+\rho_2)(V_1 + V_2) =\frac1{2}\intl_\aS\rho_1V_1 + \frac1{2}\int\rho_1V_2 + 
\frac1{2}\intl_\aS\rho_2V_1+ \frac1{2}\intl_\aS\rho_2V_2.$$

The first and the last term are the corresponding self-energies. The other two terms give the energy of interaction; note that
$$\intl_\aS\rho_1V_2 = \intl_\aS\rho_2V_1 = \frac1{4\pi}\intl_\aS\intl_\aS\frac{\rho_1(q_1)\rho_2(q_2)}{|q_1-q_2|}\ dq_1\ dq_2.$$

If one considers that $\rho_2$ acts on $\rho_1$ i.e. action is taken instead of interaction, then by
$$\frac1{2}\intl_\aS\rho_1V_2 +\frac1{2}\intl_\aS\rho_2V_1 = \intl_\aS\rho_1V_2$$
one gets the extraneous energy of $\rho_1$ in the field of $\rho_2$.

Let us continue the usual argumentation regarding self-energy. From the Maxwell equations $\nabla\cdot\E=\rho$ and $\E=-\nabla V$, 
the self-energy density can be written in the form
\beq\label{manip} \frac1{2}\rho V=\frac1{2}\bigl(\nabla\cdot(\E V) + |\E|^2\bigr).\enq

If the self-energy density is integrable and $V$ and $\E$ are sufficiently smooth and tend to zero at infinity 
in a convenient order (which is satisfied e.g. if $\rho$ has compact support), then applying Gauss' theorem 
for a ball around an arbitrary space point, one obtains that the integral of $\nabla\cdot(\E V)$ is zero, thus 

\beq\label{rov} \frac1{2}\intl_{\aS} \rho V= \frac1{2}\intl_{\aS}|\E|^2.\enq

From that, one concludes that $\frac1{2}|\E|^2$ is the electric energy density of the electric field $\E$ (\cite{Jackson} p.21).

This is an erroneous conclusion.

1. Equality \eqref{rov} is all right, provided the conditions are fulfilled. 
The integral of $\frac1{2}|\E|^2$ equals the integral of $\frac1{2}\rho V$, but the 
functions themselves are different; in particular, $\rho V$ is zero where $\rho$ is zero but $|\E|^2$ is not 
zero there. Therefore, {\it it is unjustified} to consider $\frac1{2}|\E|^2$, instead of $\frac1{2}\rho V$, the electric 
self-energy density of the charge distribution $\rho$.

2. {\it It is more unjustified} that $\frac1{2}|\E|^2$, instead of the self-energy density of the charge distribution,
is considered the energy density of the electric field; this {\it mis-switching} is due to the fact $\frac1{2}|\E|^2$ 
refers to the electric field only (does not refer to any charges).

Moreover, let us see clearly: for an electric field $\Ef$, $|\Ef|^2=\Ef\cdot\Ef$, as the product of two Distributions, 
makes no sense, in general. For a field function $\E$ the following cases can occur: 

\begin{itemize}

\item[A.] $\frac1{2}|\E|^2$ is integrable; then it defines a regular Distribution. Nevertheless, it cannot be 
considered the self-energy density of a charge distribution; {\it it has the only physical meaning} that 
its integral over all the space -- i.e. as a regular Distribution, applied 
to the constant function $1$ -- equals the total electrostatic self-energy of the charge distribution producing $\E$.

\item[B.] $\frac1{2}|\E|^2$ is not integrable but is locally integrable; then it defines a regular Distribution. 
Nevertheless, it cannot be considered a (self-)energy density and {\it it is questionable} whether it has a physical
meaning. 

\item[C.] $\frac1{2}|\E|^2$ is not locally integrable; then it does not define a Distribution without further ado.             

\end{itemize}

\subsection{Point charge, self-energy}\label{ponttame}

The `energy density' $\frac1{2}|\E|^2$ of a point charge is not integrable, so usually one states that the electric 
energy of a point charge is infinite.

{\it It is amazing and shocking that from the first step, ``the electric energy of a single point charge is zero'',
one arrives  at the conclusion ``the electric energy of a single point charge is infinite''.}

Infinite energy is a result of a three times incorrect reasoning: 
\begin{itemize}

\item  $\frac1{2}\rho V$ makes no sense for a point charge, 

\item  Gauss' theorem for $\nabla\cdot(\E V)$ cannot be applied in the case of a point charge, 

\item  $\frac1{2}|\E|^2$ is not the energy density even for a continuous charge density.
\end{itemize}

Infinite electric energy is a nonsense; there are two usual ways to eliminate it, based on the mass-energy equivalence.

1. The actual finite mass of a charged point particle is the sum of a negative infinite mechanical mass and the positive 
infinite electric mass (energy).

2. A charged particle is not point-like; then considering it a continuously charged ball, its classical radius is determined
from its known electric energy (mass).  

Then we can put the question: all these do not apply for a neutral particle; what about its mass and radius?  

Seriously speaking, let us realize that point charges, as well as continuous charge distributions, are models. Considering
a material object point-like, we do not assert that it is a point (the Earth is not a point because its orbit  
around the Sun is calculated in this way).

Let me emphasize: the nonsense of infinite electric energy of a point charge is the result of a 
three times incorrect reasoning; {\it we have to get rid of this reasoning itself}.

It can be stated unmistakably that the self-energy of a point charge is zero.

The fictitious `self-energy density'
$$\frac1{2}|\E(q)|^2=\frac1{2}\frac{e^2}{16\pi^2}\frac{1}{|q-q_o|^4}$$ 
of a point charge $e$ resting in the space point $q_o$, not being locally integrable, does not define a regular Distribution. 
It has a pole at $q_o$, and 
we can attach to it a Distribution by pole taming (see Section \ref{polet}). Introducing $\q:=q-q_o$, we have  
\beq\label{selfen}(\tm|\E|^2\mid\psi)=\intl_\Sd \E(q_o + \q)|^2(\psi(\q)-\psi(\0))\ d\q \ !!\enq
for all test functions $\psi$.                            

$\frac1{2}\tm|\E|^2$ is a mathematical object; has it some physical meaning? 
It is not a measure, and even if it were, it could not be considered the self-energy distribution (see cases A. and B.).
Its only physical meaning could be (see case A.) that its `integral' makes sense i.e. it 
can be applied to the constant function $1$ and the result is zero (the self-energy of the point charge). This is satisfied:

\begin{all} 
\beq\label{sajen}\left(\frac1{2}\tm|\E|^2\Bigm| 1\right)=0.\enq \end{all}

{\bf Proof} For a given $0<a$ let us take the series $\omega_n:=\omega_{na,(n+1)a}$ defined by \eqref{omeg}. Then we have
\begin{align*} \bigl(\tm|\E|^2\mid \omega_n\bigr) &=
\intl_{\Sd}|\E(q_o +\q)|^2 \bigl(\omega_n(\q)-\omega_n(0)\bigr)\ d\q \ !!=\\
&=4\pi\frac{e^2}{16\pi^2}\left(\intl_{na}^{(n+1)a} \frac{\omega_n(r)-1}{r^2}\ dr 
-\intl_{(n+1)a}^\infty\frac1{r^2}\ dr\right).\end{align*}

Since $|\omega_n-1|\le1$, the absolute value of the first integral above is majorized by the integral of $\frac1{r^2}$ which is 
$\frac1{na}-\frac1{(n+1)a}$; thus,
$\lim\limits_{n\to\infty}\bigl(\tm|\E|^2\mid \omega_n\bigr)=0$. $\mqed$
                                                                                                                     
As a consequence, not without any reservations, $\frac1{2}\tm|\E|^2$ can be considered 
the {\it Distribution of self-energy} (which is not a self-energy distribution!)
of a point charge, keeping in mind its only physical meaning: {\it the self-energy is zero}.

{\bf Remark} In the not too simple theory of Colombeau, 
$\frac1{2}|\Ef|^2$ as a product of Distributions can be defined, and it is obtained (see \cite{Gsproner}) 
that the energy is infinite but is located at the position of the point charge rather than  spread over the whole space;
this is not a too reasonable result because of the infinity.

\subsection{Self-energies, charges and dipoles}\label{sendip}

For point dipoles, we can copy the formulae obtained for charges.

The self-energy of a point dipole is zero.
 
The self-energy distribution of the dipole distribution ($n\ge2$)
\beq\label{dip} \sum_{k=1}^n\pp_k\delta_{q_k}\enq
is
\beq\label{dipelo} -\frac1{2}\sum_{k=1}^n\pp_k\cdot(\E-\E_k)\delta_{q_k};\enq
its integral is the total self-energy
\beq -\frac1{2}\sum_{k=1}^n\pp_k\cdot(\E-\E_k)(q_k).\enq

Note that $\pp_k\cdot(\E-\E_k)\delta_{q_k}$ is correct because $\E-\E_k$ is the sum of the fields of the charges except the $k$-th one. 
$\E_k$ is not defined at $q_k$, thus $\E_k\delta_{q_k}$ makes no sense according to the original mathematical definition 
of a measure multiplied by a function. The fact that the electric self-energy of a single 
dipole is zero suggests that we define formally $\pp_k\cdot\E_k\delta_{q_k}:=0$. Then $\E_k$ can be omitted from \eqref{dipelo}.

Putting together more and more, smaller and smaller dipoles, one goes over -- not in a mathematically 
exact sense -- to a continuous dipole distribution, thus, without a serious objection, 
$$-\frac1{2}\Pt\cdot\E$$ 
{\it can be accepted as the  self-energy density of a static dipole density $\Pt$ in its own field} $\E$, provided that 
$\Pt\cdot\E$ is locally integrable (and even integrable)

Let us now consider a system consisting of point charges and point dipoles described by the distribution 
\beq\label{chardip} \sum_{k=1}^m e_k\delta_{q_k} + \sum_{k=m+1}^n\pp_k\delta_{q_k}.\enq

The energy of the charge $e_i$ in the potential $V_j$ of the dipole $\pp_j$ is $e_iV_j(q_i)$; 
the energy of the  dipole $\pp_j$ in the electric field $\E_i$ of the charge $e_i$ is $-\pp_j\cdot\E_i(q_j)$.
According to the formulae of potentials and fields, $e_iV_j(q_i)=-\pp_j\cdot\E_i(q_j)$. 
Therefore, the self-energy of this system of charges and dipoles is
\begin{multline}\label{diponten} \frac1{2}\sum_{k=1}^me_k\left(\sum_{k\neq i=1}^mV_i(q_k) + \sum_{j=m+1}^nV_j(q_k)\right)-\\
-\frac1{2}\sum_{j=m+1}^n\pp_j\cdot\left(\sum_{i=1}^m\E_i(q_j) + \sum_{j\neq l=m+1}^n\E_l(q_j)\right).\end{multline}
Then, as previously, {\it it can be accepted that 
$$\frac1{2}(\rho V - \Pt\cdot\E)$$
is the self-energy density of the charge density  $\rho$ and the dipole density $\Pt$ in their own fields}, provided that
it is locally integrable (and even integrable).

Now two manipulations, similar to the one resulting in \eqref{manip}, can be made. From the Maxwell equations 
$\nabla\cdot\E=\rho -\nabla\cdot\Pt$ and $\E=-\nabla V$, one has

1. $\rho V= \nabla\cdot(\E+\Pt) V= \nabla\cdot((\E+\Pt)V) + (\E + \Pt)\cdot\E$; 
the integral of the divergence (with convenient integrability and differentiability conditions) over all the space will be zero,
thus, the total self-energy is 
\beq\label{manipd}\frac1{2}\intl_\aS|\E|^2.\enq

2. $-\Pt\cdot\E=\Pt\cdot\nabla V=\nabla\cdot(\Pt V) - (\nabla\cdot\Pt)V = \nabla\cdot(\Pt V) + (\nabla\cdot\E -\rho)V =
\nabla\cdot((\Pt +\E)V) - \rho V + |\E|^2$; integrating over all the space, the same result \eqref{manipd} is obtained.

Then one can have the erroneous conclusion that \eqref{manipd} is the energy density of the electric field regardless
of its source and we can repeat, according to the sense, what we said at the and of Subsection \ref{sselfen}.  

As concerns a single point dipole, it can be stated unmistakably that its self-energy is zero.
The square of \eqref{dipfu} is not locally integrable having a pole; taming it  
and applying $\frac1{2}\tm|\E|^2$ to the constant function $1$ we get the corresponding self-energy.

{\bf Remark} Note that in the general case of charges and dipoles, \eqref{manipd} is obtained by omitting terms with divergence
as it is usual for charges. Who accepts this manipulation for charges, must accept it 
for the general case, too, and must arrive at $\frac1{2}|\E|^2$, which contradicts the other usual formula $\frac1{2}\E\cdot\Dt$, 
obtained by a method far from being exact (\cite{Jackson} p.166).

\section{About forces in electrostatics}\label{force}

\subsection{Extraneous forces}

The electric force acting on a point charge $e$ resting at the space point $q_o$ in an {\it extraneous} static electric 
field $\E$ is $e\E(q_o)$, provided that $\E$ is defined at $q_o$. 

From that, it is obvious that the $\rho\E$ can be accepted as the static electric force density 
acting on a charge density $\rho$ in an {\it extraneous} static electric field  $\E$, provided that $\rho\E$ is locally
integrable i.e. it defines a regular Distribution.

The electric force acting on a point dipole  $\pp$ at the space point 
$q_o$ in an {\it extraneous} static electric field $\E$ is $\pp\cdot\nabla\E(q_o)=(\nabla\E(q_o))\cdot\pp$ (\cite{Jackson}), 
provided that $\E$ is defined at $q_o$.
Then by Maxwell equation $\nabla\times\E=0$, \ $\Pt\cdot\nabla\E=(\nabla\E)\cdot\Pt$ can be accepted as the 
static electric force density 
acting on a dipole density $\Pt$ in an {\it extraneous} static electric field $\E$, provided that 
$\Pt\cdot\nabla\E=(\nabla\E)\cdot\Pt$ is locally integrable.

The notion of forces in an extraneous field is a consequence of the fact that only an 
action is taken into account instead of an interaction.

\subsection{Self-forces, charges}

An inertial point charge does not act on itself: its self-force is zero.

The electric self-force distribution of the charge distribution in \eqref{pontelo} is
\beq\label{eroel}\sum_{k=1}^ne_k(\E-\E_k)\delta_{q_k}\enq
where $\E$ is the field produced by all the charges and $\E_k$ is the field produced by the 
$k$-th charge. 

$e_k(\E-\E_k)\delta_{q_k}$ is correct as previously, and the fact that an inertial point charge does not act on itself  
suggests that we define formally $e_k\E_k\delta_{q_k}:=0$. Then $\E_k$ can be omitted from \eqref{eroel}.

Putting together more and more particles with smaller and smaller charges, one goes over to a continuous charge distribution, thus,
without a serious objection, 
\beq\label{conter}\rho\E\enq
{\it can be accepted as the static electric self-force density of the charge density $\rho$ in its own electric field} $\E$, 
provided that $\rho\E$ is locally integrable.
                                                                          
{\bf Remarks} 1. The formula  of self-force density equals the formula of extraneous force density: both are $\rho\E$ 
with different physical meaning. This warns us that it is worth keeping in mind the physical meaning
of an actual formula, not to be confused.

2. Both in the discrete case and in the continuous case, some other  non-electric 
(e.g. mechanical) force distribution, opposite to the electric one, 
must be applied in order that the charges do not move from their given places, 

3. If $\rho\E$ is integrable, then the total electric force is zero  (as it must be for Newton's law of action-reaction): 
$\rho\E=(\nabla\cdot\E)\E= \frac1{2}\nabla\cdot(|\E|^2\1)$ where $\1$ is the identity of $\Sd$; the integral of the 
divergence over all the space is zero.

\subsection{`Stress tensor', charges}\label{strch}

As usual, one introduces 
\beq\label{stress}\Ll:=-\E\otimes\E + \frac1{2}|\E|^2\1\enq
($\1$ is the identity map of $\Sd$) for an electric field function $\E$  
and using the Maxwell equation $\nabla\cdot\E=\rho$  one has
\beq\label{negdiv}-\nabla\cdot\Ll= \rho\E;\enq
here $\rho\E$ is the self-force density of the charge density $\rho$ because $\E$ is produced by $\rho$.

On the analogy of continuum mechanics where the force density is the negative divergence of a stress tensor
one considers $\Ll$ the stress tensor of the electric field $\E$. 
 
There are mistakes in this reasoning.

1. Equality \eqref{negdiv} is all right, provided the conditions are fulfilled. 
In continuum mechanics, the stress tensor has a real physical meaning by giving the shearing forces etc. inside the continuum. 
There the stress tensor is concentrated to the body of the 
continuum, the stress tensor is zero where there is no material. Here only the negative divergence of $\Ll$ has a physical meaning
and $\Ll$ is not zero where $\rho$ is zero, thus, {\it it is unjustified} to consider $\Ll$ the electric self-stress tensor 
of the charge distribution $\rho$. 

2. {\it It is more unjustified} that $\Ll$, instead of the self-stress tensor of the charge distribution, is 
considered the stress tensor of the electric field; this {\it mis-switching} is due to the fact that $\Ll$ refers to the 
electric field only (does not refer to any charges).

Moreover, let us see clearly: for an electric field $\Ef$, $\Ef\otimes\Ef$ and $|\Ef|^2=\Ef\cdot\Ef$, as the products 
of two Distributions, make no sense, in general. For a field function $\E$ the following cases can occur: 

\begin{itemize}

\item[A.] $\Ll$ and $\nabla\cdot\Ll$ are locally integrable; then 
$\Ll$ defines a regular Distribution, nevertheless, it cannot be considered a stress tensor, and 
{\it it has the only physical meaning} that its negative divergence is the self-force density. 

\item[B.]  $\Ll$ is locally integrable but $\nabla\cdot\Ll$ is not; then $\Ll$ defines a regular Distribution 
but {\it it is questionable} whether it has some physical meaning,, 

\item[C.] $\Ll$ is not locally integrable; then it does not define a Distribution without further ado.

\end{itemize}

\subsection{`Stress tensor', point charge}

For a point charge the fictitious `stress tensor' 
\beq\label{selfstr}\Ll(q)=\frac{e^2}{16\pi^2}\left(-\frac{(q-q_o)\otimes(q-q_o)}{|q -q_o|^6} + \frac{\1}{2|q-q_o|^4}\right)\enq
is not locally integrable, hence it makes no a priori sense as a Distribution. It has a pole at $q_o$, thus, as for the self-energy, 
we can attach to it a Distribution by pole taming. With the notations $\q:=q-q_o$ and $\nb(\q):=\frac{\q}{|\q|}$,
\begin{align}\label{pontfesz}\Ll(q_o +\q)&=\frac{e^2}{16\pi^2}\left(\frac{-\nb(\q)\otimes\nb(\q)}{|\q|^4} + 
\frac{\1}{2|\q|^4}\right)= \\ &=\frac{e^2}{16\pi^2}\frac{\1-2\nb(\q)\otimes\nb(\q)}{2|\q|^4}.\end{align}

The numerator is an even tensor power of $\nb$, therefore 
$$(\tm\Ll\mid\psi)=\intl_\Sd \Ll(q_o+\q)\bigl(\psi(\q)-\psi(\0)\bigr)\ d\q\ !!$$
holds for all test functions $\psi$; by the equality (see \eqref{nn})
$$\intl_{S_1(\0)}(\1-2\nb\otimes\nb)\ d\nb= \frac{4\pi}{3}\1,$$
it can be rewritten in the form
$$(\tm\Ll\mid\psi)=\intl_\Sd \left(\Ll(q_o +\q)\psi(\q) - 
\frac{e^2}{16\pi^2}\frac{\psi(\0)}{6|\q|^4}\1\right)\ d\q.\ !!$$

$\tm\Ll$ is a mathematical object; has it some physical meaning? 
It is not a measure, and even if it were, it could not be considered the stress distribution.
Its only physical meaning could be that its negative divergence is zero (the self-force of the point charge). 
This is satisfied:

\begin{all} \beq\label{nabp}-\nabla\cdot\tm\Ll =0.\enq \end{all}

{\bf Proof} The divergence of the function $\Ll$ (not defined at $q_o$) is zero: there is no force outside the charge, but 
this does not mean that the divergence of $\tm\Ll$ is zero as well. According to the definition of the derivatives of Distributions,
\begin{align}(-\nabla\cdot\tm\Ll\mid\psi)&=(\tm\Ll\mid\cdot\nabla\psi)= \\
&=\intl_\Sd \left(\Ll(q_e+\q)\cdot\nabla\psi(\q) - \frac{e^2}{16\pi^2}\frac{\nabla\psi(\0)\1}{6|\q|^4}\right)\ d\q.\ !!
\end{align}
 
The first term in the integral is $\nabla\cdot(\Ll\psi)$ because $\nabla\cdot\Ll=0$. 
The second term, with the notation $r(\q):=|\q|$ and
\beq \nabla\cdot\frac{\nb}{r^3}=-\frac1{r^4},\enq
can be written in the form
$\nabla\cdot\left(\frac{e^2}{16\pi^2}\frac{(\nabla\psi(\0))\otimes\nb}{6r^3}\right)$.

Then taking $\intl_\Sd=\lim\limits_{R\to0}\intl_{r>R}$ and applying Gauss' theorem,
we obtain
\begin{multline*}(-\nabla\cdot\tm\Ll\mid\psi) =\\ 
=\frac{e^2}{16\pi^2}\lim_{R\to0}\intl_{S_1(\0)}\left(\frac{\1 - 2\nb\otimes\nb}{2R^2}\psi(R\nb) 
+ \frac{(\nabla\psi(\0))\otimes\nb}{6R}\right)\cdot(-\nb) \ d\nb.\end{multline*}
\medskip

The integral of the second term is
$$\intl_{S_1(\0)}\frac{(\nabla\psi(\0))\otimes\nb}{6R}\cdot(-\nb) \ d\nb=-\frac{4\pi}{6R}\nabla\psi(\0).$$

As concerns the first term, we take the expansion 
$$\psi(R\nb)=\psi(\0) + \nabla\psi(\0)\cdot(R\nb) + \frac1{2}(R\nb)\cdot\nabla^2\psi(\0)\cdot(R\nb) + \text{ordo}(R^2).$$

1. The integral of $\text{ordo}(R^2)$ tends to zero as $R$ tends to zero.
                                                    
2. The integrals with $\psi(\0)$ and $\frac1{2}(R\nb)\cdot\nabla^2\psi(\0)\cdot(R\nb)$ are zero 
because of $(\1 - 2\nb\otimes\nb)\cdot(-\nb)=\nb$ and $\eqref{n1}$,

3. The integral with $\nabla\psi(\0)\cdot(R\nb)$, because of \eqref{nn}, becomes

$$\intl_{S_1(\0)}\bigl(\nabla\psi(\0)\cdot(R\nb)\bigr)\frac{\nb}{2R^2}\ d\nb=\frac{4\pi}{6R}\nabla\psi(\0). \mqed $$

As  a consequence, not without any reservations, $\tm\Ll$ can be considered 
the  {\it Distribution of self-stress} (which 
is not a self-stress distribution!) of a point charge, keeping in mind
that only its negative divergence has a physical meaning: the self-force is zero.

\subsection{Self-forces, charges and dipoles}\label{dipfor}

An inertial point dipole does not act on itself, its self-force is zero.

The self-force distribution of the dipole distribution \eqref{dip} is
\beq \sum_{k=1}^n\pp_k\cdot\nabla(\E-\E_k)\delta_{q_k}.\enq

Then 
$$\Pt\cdot\nabla\E=(\nabla\E)\cdot\Pt$$
can be accepted as the electric self-force density of the dipole density $\Pt$ in its own electric field $\E$, 
provided that $\Pt\cdot\nabla\E=(\nabla\E)\cdot\Pt$ is locally integrable.

Let us now consider a system consisting of point charges and point dipoles having the distribution \eqref{chardip}.

The force acting on the charge $e_i$ in the field $\E_j$ of the dipole $\pp_j$ is $e_i\E_j(q_i)$; 
the force acting on the  dipole $\pp_j$ in the electric field $\E_i$ of the charge $e_i$ is $\pp_j\cdot(\nabla\E_i)(q_j)$
and we have $e_i\E_j(q_i)=-\pp_j\cdot(\nabla\E_i)(q_j)$. 
Then by an equality similar to \eqref{diponten}, {\it it can be accepted that 
$$\rho \E + \Pt\cdot\nabla\E$$
is the self-force density of the charge density  $\rho$ and the dipole density $\Pt$ in their own fields}, provided that 
it is locally integrable.

Now the fictitious `stress tensor'
\beq\label{stressp}\Ll:=-\E\otimes\E + \frac1{2}|\E|^2\1-\E\otm\Pt=-\E\otimes\Dt + \frac1{2}|\E|^2\1\enq
has the property that its negative divergence is the formal self-force density
\beq -\nabla\cdot\Ll=\rho\E +\Pt\cdot\nabla\E.\enq

Then we can repeat the problems listed at the end of Subsection \ref{strch}, crowned by the fact that $\Ll$ 
is not given by field quantities only (contains $\Pt$ explicitely or hidden in $\Dt$).

For a point dipole, the fictitious `stress tensor' is questionable: $\E\otm\pp\delta_{q_o}$ makes no sense.
This term omitted, $\Ll$ becomes a function having a pole; whoever wants, tame it and calculate the divergence of $\tm\Ll$. 

\section{A summary}\label{summa}

Let us survey the ways we followed treating self-energies and self-forces in statics. 

1. The starting points were zero self-energy and zero self-force of point charges.

2. Formulae \eqref{ponten} and \eqref{eroel} of self-energy and self-force of 
a system of point charges suggested  formulae \eqref{conten} and \eqref{conter} 
of self-energy and self-force density of a continuous charge distribution.

3. a. The formula of self-energy density is usually transformed by \eqref{rov} into the false form $\frac1{2}|\E|^2$ which 
can have the only real physical meaning that its integral is the total self-energy of the 
continuous charge distribution.

3. b. The usual interpretation of \eqref{stress} as a stress tensor is not right; it can have the only real 
physical meaning that its negative divergence is the self-force density of the continuous charge distribution.
 
4. The fictitious `self-energy density' and `stress tensor' of a point charge are not locally integrable; 
we attached to them Distributions by pole taming.

5. Those Distributions have the correct physical meaning: the total self-energy and the self-force of a point charge 
are zero.

We have made a detour from a point charge to charge densities and back, by giving sense to originally doubtful 
or unjustified formulae. This going round and arriving back at the starting point will be useful for 
later considerations.

For dipoles, steps 1. - 3. are similar but steps 4. and 5. are questionable.

\section{Beyond statics \\ with respect to a standard inertial frame}\label{beyond}            

\subsection{Basic notions}

For an {\bf arbitrary} standard inertial frame $\uu$ the electric and magnetic quantities are the following:

\begin{align*}
\Ef:& \ \uu\text{-electric field},\quad \Bf: \ \uu\text{-magnetic field},\\
\e:& \ \uu\text{-charge distribution},\quad \ij: \ \uu\text{-current distribution},\\
\Pf:& \ \uu\text{-electric dipole distribution},\quad \Mf:\ \uu\text{-magnetic moment distribution},\end{align*}
each of them depends on $\uu$-time and $\uu$-space.

The Maxwell equations are
\beq\label{mxw1} \nabla\cdot\Ef=\e-\nabla\cdot\Pf, \qquad -\pt_t\Ef +\nabla\times\Bf= \ij + \pt_t\Pf +\nabla\times\Mf,\enq
\beq\label{mxw2} \nabla\cdot\Bf= 0, \qquad \pt_t\Bf +\nabla\times\Ef=0. \enq

A special case is when the charge etc. distributions have continuously differentiable densities $\rho$, $\ii$, $\Pt$ and $\Mt$, 
respectively, and the fields are regular Distribution corespontding to continouosly differentiable  functions $\E$ and $\B$; 
then the Maxwell equations are
\beq\label{mxw11} \nabla\cdot\E=\rho-\nabla\cdot\Pt, \qquad -\pt_t\E +\nabla\times\B= \ii + \pt_t\Pt +\nabla\times\Mt,\enq
\beq\label{mxw22} \nabla\cdot\B= 0, \qquad \pt_t\B +\nabla\times\E=0. \enq

Further, with the quantities 
\beq\label{debh}\Dt:=\E+\Pt, \qquad \Ht:=\B-\Mt,\enq
\eqref{mxw11} gets its most frequently used form
\beq\label{mxwdh} \nabla\cdot\Dt=\rho, \qquad -\pt_t\Dt +\nabla\times\Ht= \ii.\enq
This simpler form, however, is misleading because $\Dt$ and $\Ht$ are considered proper field quantities though they are not 
because of including dipoles and magnets hidden. (And even, from the start of the theory of electromagnetism, 
for a long time, the roles of $\Ht$ and $\B$ have been confused.) 

Two charge etc. distributions  are {\bf extraneous} to each other if their supports are disjoint at every instant. 
A charge etc. distribution and an electromagnetic field are called {\bf extraneous} to each other according to the sense.

The {\bf self-field} of a charge etc. distribution is the electromagnetic field produced by them.

In the following we assume that convenient conditions are fulfilled in order that the formulae make sense as Distributions, so 
``provided that ...'' will be omitted.                                                         

\subsection{Extraneous forces, charges}

Let us consider a point charge $e$ moving under the action of an electric and magnetic field 
$\E$ and $\B$, respectively, its given motion being desribed by the function $t\mapsto\rs(t)$. Then the {\it extraneous force} acting 
on the particle is the well-known Lorentz force $e(\E + \rs'\times\B)$, $\rs'$ being the velocity of the point charge.

From that it is quite obvious that
\beq\label{lorero}\rho(\E+ \vv\times\B)=\rho\E + \ii\times\B\enq
can be accepted as the electromagnetic force density  
acting on a charge density $\rho$, with velocity field $\vv$ (and $\ii:=\rho\vv$), in an {\it extraneous} electric and magnetic 
field $\E$ and $\B$.

Further, during the motion of a point charge, the mechanical power of the extraneous fields is $e\E\cdot\rs'$. Consequently,  
\beq\label{telj}\rho\E\cdot\vv=\E\cdot\ii\enq
can be accepted as the mechanical power density in an {\it extraneous} field $\E$ acting on the current density $\ii$.

\subsection{Self-forces, charges}\label{quest}

Let us try to follow the route  suggested by statics according to Subsection \ref{summa}.

Then there is a problem at the first step: a non-inertial point charge  acts on itself depending on how it moves, 
but its self-force, denoted by $\f_s$, is not known at present. Hopefully, we get it at the end of the route.

Let us consider a system of point charges $e_1,\dots,e_n$ with given motions $\rs_1,\dots\rs_n$, respectively; then
the self-force acting in this system, with a straightforward (and somewhat loose) notation, is
\beq\label{nisajer}\sum_{k=1}^ne_k\bigl((\E-\E_k) + \rs_k'\times(\B-\B_k)\bigr) + \sum_{k=1}^n\f_{s,k}\enq
where $\E$ and $\B$ are the fields produced by all the charges, $\E_k$ and $\B_k$ are the fields produced 
by the $k$-th charge.

As in statics, $(\E-\E_k)(t,\rs_k(t))$ and $(\B-\B_k)(t,\rs_k(t))$ are correct.
\medskip

If, analogously to statics, we defined the originally senseless $e_k(\E_k + \rs_k'\times\B_k)$ being 
formally equal to the unknown self force $\f_{s,k}$ then $\E_k$ and $\B_k$ as well as $\sum_k\f_{s,k}$ would disappear 
from the formula above. But now nothing authorizes us to do so because, contrary to statics, now the self-force is not
known. 

Putting together more and more particles with smaller and smaller charges, we go over a continuous charge distribution,
obtaining the self-force density
\beq\label{lorplus}\rho(\E + \vv\times\B)\ +\ \boldsymbol?\enq
where $\E$ and $\B$ are the fields produced by the charge distribution, $\vv$ is a velocity field and $\boldsymbol?$ is something. 
Note that the first term in \eqref{lorplus} 
equals \eqref{lorero} having a different physical meanings, similarly to the static case; but the whole \eqref{lorplus}
differs possibly from \eqref{lorero}.

Thus, contrary to electrostatics, {\it it is questionable} whether $\rho(\E + \vv\times\B)=\rho\E + \ii\times\B$ 
is the self-force density of $\rho$ in its own fields $\E$ and $\B$.
                                   
The self-power density in the system of charges is
\beq\label{nisajt}\sum_{k=1}^ne_k(\E-\E_k)\cdot\rs_k' + \sum_{k=1}^n\f_{s,k}\cdot\rs_k'\enq
and going over to a continuous charge distribution, we get 
\beq\label{selfpw}\E\cdot\ii\ +\ ?.\enq

Thus, {\it it is questionable} whether $\E\cdot\ii$ is the self-power density of $\ii$ in its own field $\E$.

\subsection{Balance equations? 1}\label{bal1}

Let us see a usual argumentation (see e.g. \cite{Groot}). 

Let the electric and magnetic field  $\E$ and $\B$ be produced by the charge density $\rho$ 
and current density $\ii$ (electric dipoles and magnetic moments are not present).
Subtracting the scalar product of the second equation  
in \eqref{mxw22} by $\B$ from the scalar product of the second equation in \eqref{mxw11} by $\E$ one gets

\beq\label{poyn}-\pt_t\frac1{2}(|\E|^2 + |\B|^2)-\nabla\cdot(\E\times\B) = \E\cdot\ii\enq
which is considered an energy balance equation, with 
\begin{itemize}
\item the power density 
$$\E\cdot\ii,$$
\item the energy flow density  (Poynting vector) 
\beq\label{enaram}\E\times\B,\enq
\item the energy density 
\beq\label{endens}\frac1{2}(|\E|^2 + |\B|^2).\enq
\end{itemize}

There are some serious doubts, however. 

1. Since the fields in the Maxwell equations are just the ones produced by the charges etc., $\E\cdot\ii$ 
should be the self-power density which is questionable according to \eqref{selfpw}.

2. The doubts regarding the energy flow density are well-known (\cite {Feynman}): first, it is 
unique only up to a divergence-free term; second, it is not zero for a 
static $\E$ and a static $\B$  when nothing happens, nothing moves with respect to the reference frame 
in question; does the energy really flow in that reference frame?

3. According to Subsection \ref{sselfen}, it is unjustified to take \eqref{endens} 
either for self-energy or energy density. By the way, which is not mentioned usually, the presumed energy density 
is unique only up to a time-independent term. 
\medskip

Further, adding the vectorial product of the second equation in \eqref{mxw11} by $\B$ to  
the vectorial product of the second equation in \eqref{mxw22} by $\E$, using the other two 
equations, too, one gets

\beq\label{poynt}-\pt_t(\E\times\B) - \nabla\cdot\left(-\E\otimes\E - \B\otimes\B + \frac1{2}(|\E|^2 + |\B|^2)\ \1\right) 
= \rho\E + \ii\times\B\enq                                           
which is considered a momentum balance equation, with
\begin{itemize}
\item the force density 
$$\rho\E + \ii\times\B,$$
\item the momentum flow density  
\beq\label{streb}-\E\otimes\E - \B\otimes\B + \frac1{2}(|\E|^2 + |\B|^2)\ \1,\enq
\item the momentum density 
$$\E\times\B.$$
\end{itemize}

There are some serious doubts, however. 

1. Since the fields in the Maxwell equations are just the ones produced by the charges etc., $\rho\E + \ii\times\B$ 
should be the self-force density which is questionable according to \eqref{lorplus}.

2. According to Subsection \ref{strch}, it is unjustified to take \eqref{streb} 
for the momentum flow density (stress tensor). 
                     
3. The Poynting vector can hardly be a momentum density for static fields when nothing moves.

{\bf Remark} In spite of the listed doubts, it is not excluded that {\it beyond statics, in some circumstances} 
\eqref{endens} is the energy density, \eqref{enaram} is the energy flow density etc. 

{\it The problem is how those  `some circumstances' can be precisely formulated; it is 
sure only that they are connected with radiation because it is an everyday experience that energy flows with light}.

\subsection{Balance equations? 2}\label{bal2}

If there are an electric dipole density $\Pt$ and 
magnetic moment density $\Mt$, too, then instead of \eqref{poyn} one gets
\beq\label{ebpm}-\pt_t\frac1{2}(|\E|^2 + |\B|^2)-\nabla\cdot(\E\times\B) = \E\cdot\ii + \E\cdot\pt_t\Pt -\B\cdot\pt_t\Mt + 
\nabla\cdot(\E\times\Mt)\enq
which has not the form of a balance equation. 

It becomes simpler with the quantities \eqref{debh}, without becoming a balance equation: 
$$-\bigl(\E\cdot\pt_t\Dt + \Ht\cdot\pt_t\B\bigr)-\nabla\cdot(\E\times\Ht)=\E\cdot\ii$$

To obtain a balance-like equation at all costs, one considers (\cite{Jackson} p.189) the special case when 
there are no independent dipoles and magnets, the whole space of a given inertial frame is filled 
with a homogeneous material which is polarized and magnetized i.e. $\Dt=\epsilon\E$ and $\mu\Ht=\B$ holds; then 
\beq\label{edbh}-\pt_t\frac1{2}(\E\cdot\Dt + \B\cdot\Ht) -\nabla\cdot(\E\times\Ht)=\E\cdot\ii.\enq
This is a balance-like equation, but in vain, more serious doubts arise as previously.

The situation is even worse for the general momentum balance: then the formula obtained instead of \eqref{poynt}
cannot be converted into a balance-like equation for momentum even in the special case above.

{\bf Remarks} 1. There are some other forms of energy and momentum balance (see \cite{Jackson} and a good review in 
\cite{JimCamLop}); all of them are unsatisfactory in some aspect, and all of them have the common fault that 
$\frac1{2}(|\E|^2 + |\B|^2)$ or $\frac1{2}(\E\cdot\Dt + \B\cdot\Ht)$ or similar formulae are accepted as energy density.

2. The not too convincing method in statics to obtain the energy density $\frac1{2}\E\cdot\Dt$ (see the Remark at the end of
Subsection \ref{sendip}) is probably invented for an acceptable explanation of the balance equation \eqref{edbh}. 

3. The problems connected with static dipole distributions, outlined in Subsections \ref{sendip} and \ref{dipfor},
are precursors of the more significant problems that emerged in balance equations. 

\section{Spacetime formulation}\label{spteld}

\subsection{Basic notions}

As it is said in the Intoduction, coordinet-free spacetime notations (Lorentz product, differentiation etc.) 
will be used according to \cite{Matolcsi} and \cite{Mato}.

Electric and magnetic fields, $\Ef$ and $\Bf$, according to a standard inertial frame, are split components 
of an electromagnetic field $\Ff$ which is an antisymmetric spacetime tensor Distribution.
Precisely speaking, up to now, the magnetic field has been considered, as usual, a three-dimensional vector Distribution,
but the spacelike component of $\Ff$ is a three-dimensional antisymmetric tensor which can be identified with 
a three dimensional vector.

The charge distribution and current distribution, $\e$ and $\ij$, according to a standard inertial frame, are split 
components of an absolute current $\jj$ which is a four dimensional vector Distribution. 

The electric dipole distribution $\Pf$ and magnetic moment distribution $\Mf$, according to a standard inertial frame, are split 
components of an absolute dipole-moment $\Nf$ which is an antisymmetric spacetime Distribution. 

The Maxwell equations  are

 $$\Dd\cdot\Ff=\jj +\Dd\cdot\Nf, \qquad \Dd\land\Ff=0.$$
                                                                 
With $\Gf:=\Ff-\Nf$ (whose split components according to a standard inertial 
frame are $\Df$ and $\Hf$), the first equation becomes 
$$\Dd\cdot\Gf=\jj;$$
it is advisable to keep away from it because $\Gf$ is not a proper field quantity.

A special case is when the absolute current has a continuous density $\ji$, the absolute dipole-moment has a 
continuously differentiable density $\Nt$, and the field is the regular Distribution corresponding to a 
continuously differentiable function $\F$;  then the Maxwell equations are 
$$\Dd\cdot\F=\ji +\Dd\cdot\Nt, \qquad \Dd\land\F=0.$$

Then \eqref{lorplus} and \eqref{selfpw} are the split components of 
\beq\label{absfor}\F\cdot\ji\ +\ \boldsymbol?;\enq
thus, {\it it is questionable} that $\F\cdot\ji$ is the absolute self-force density acting on $\ji$ in its own 
electromagnetic field $\F$.

\subsection{`Energy-momentum tensor'}\label{enmomt}

For lack of electromagnetic dipole-moment, the `energy density' \eqref{endens}, the `energy flow' \eqref{enaram} and 
the `stress tensor' \eqref{streb} are split components, according to a standard inertial frame, of the `energy-momentum tensor'
\beq\label{enmom}\T:= -\F\cdot\F - \frac1{4}\mathrm{Tr}(\F\cdot\F)\1;\enq
here $\1$ is the identity map of spacetime vectors.

Using the Maxwell equation $\Dd\cdot\F=\ji$ we find that
\beq\label{dfj}-\Dd\cdot\T=\F\cdot\ji,\enq                                                               
whose split components are the `balance equations' \eqref{poyn} and \eqref{poynt}.

According to the doubts emerged in Subsection \ref{bal1}, indicated by the quotation mark, 
{\it it is questionable}, in general, that \eqref{dfj} is really an energy-momentum balance.

Correspondingly, even if $\F\cdot\ji$ is the absolute self-force density (which is questionable), 
{\it it is not right} to consider \eqref{enmom}, {\it except in some circumstances}, a proper energy-momentum tensor 
(which is indicated by the quotation mark in the title). In particular, in statics, it is definitely not right.

{\it The problem is, how those `some circumstances' can be precisely formulated; it is sure only that they are connected 
with radiation.}
 
If electric dipole and magnetic moment are present, \eqref{ebpm} can be transformed into a balance-like equation \eqref{edbh}
only in exceptional cases; nevertheless, it is not a real balance equation of energy. 
Moreover, no manipulation results in a balance-like equation for momentum.

This means that it is questionable whether a general `energy-momentum tensor' exists whose negative spacetime divergence gives a 
reasonable energy-momentum balance. A number of proposals followed the first ones given 
by Abraham and Minkowski, as it is pointed out in the Introduction. The proposed different energy-momentum balances 
correspond to different physical situations (different kinds of materials and their thermodynamical properties), and only
experiments could verify or refute them.
 
\subsection{Point charge, self-force}

Let us compare our present considerations with those in statics, see Section \ref{summa}.

1. The starting point: a non-inertial point charge acts on itself, the self-force is unknown,

2. The formulae of \eqref{nisajer} and \eqref{nisajt} of self-force and self-power of a system of point charges 
suggested the formulae \eqref{lorplus} and \eqref{selfpw} 
of self-force and self-power density of a continuous charge distribution.

3. The false  `energy density' \eqref{endens}, the questionable `energy flow' \eqref{enaram} and 
unjustified `stress tensor' \eqref{streb} make questionable the balance equations \eqref{poyn} and \eqref{poynt}.

2'. In spacetime formulation, the absolute self-force density is \eqref{absfor}.

3'. It is not right to consider \eqref{enmom} a proper energy-moment tensor, in general; moreover, it is questionable 
that its negative spacetime divergence \eqref{dfj} is the absolute self-force density.
                                        
The fictitious `self-energy density' and `stress tensor' which appeared in statics now are included in the fictitious
`energy-momentum tensor'. For a static point charge, those fictitious quantities are not locally integrable in space;
Distributions can be attached to them by pole taming and those Distributions give back the starting points, the
zero self-energy and the zero self-force.

It is proved in the paper \cite{Mato} that 

4. the `fictitious energy-momentum tensor' \eqref{enmom} for a point charge with a given world line is not locally 
integrable in spacetime, and a Distribution can be attached to it by pole taming,

5. the negative spacetime divergence of that Distribution is just the well-known radiation reaction force.

Thus, we obtained what was unknown in the first step. Moreover, this result entitles us to accept that $+\ ?$ 
in \eqref{absfor} (\eqref{lorplus},\eqref{selfpw}) can be omitted.

\section{Radiation}

\subsection{Plane waves}

Elementary experimental fact is that light has wave properties and transports energy. Light far from its source 
can be modelled approximately by plane wave solutions of the Maxwell equations. The assumption that in a standard inertial frame 
$\frac1{2}(|\E|^2 + |\B|^2)$ is the energy density and  $\E\times\B$ is the energy flow density of light agrees with 
experimental data.

\subsection{Point charge}

Let us consider a point charge $e$  with a given world line function $r$. 
For a spacetime point  $x$, $\s_r(x)$ denotes the retarded proper time of $r$. Then
with the notations (\cite{Mato})                                
\beq\label{retmeny} \R_r(x):= x - r(\s_r(x)), \qquad \uu_r(x):=\dot r(\s_r(x)), \qquad \ad_r(x):=\ddot r(\s_r(x)),\enq

\beq\label{lld} \Ll_r:=\frac{\R_r}{-\uu_r\cdot\R_r}=-\Dd\s_r, \qquad \dd_r:=\ad_r + (\ad_r\cdot\Ll_r)\uu_r,\enq
the electromagnetic field produced by the point charge is the regular Distribution defined by the locally 
integrable function
\beq\label{field} \F:=\frac{e}{4\pi}\frac{\uu_r\land\Ll_r}{(-\uu_r\cdot\R_r)^2} + 
\frac{e}{4\pi}\frac{\dd_r\land\Ll_r}{-\uu_r\cdot\R_r};\enq
the first term is the {\it tied field} $\F^{td}$ which is never zero, the second term is the 
{\it radiated field} $\F^{rd}$ which is zero if and only if the charge is not accelerated.

Then the fictitious `energy-momentum tensor' \eqref{enmom} is of the form
\beq\label{enimpt}\T=\T^{td} + \T^{rtd} + \T^{rd};\enq
only the last term, the purely radiated part,
\beq\label{ezaz}\T^{rd}:= -\F^{rd}\cdot\F^{rd} - \frac1{4}\mathrm{Tr}(\F^{rd}\cdot\F^{rd})\1\enq
is locally integrable in spacetime.

It is known (\cite{Jackson} p.269) that the electric field and the magnetic field in $\F^{rd}$ have properties, similar to 
the properties of the fields in a plane wave. Then it seems natural the conjecture that \eqref{ezaz} is a real energy-momentum 
tensor in the sense that its split components \eqref{endens}, \eqref{enaram} and \eqref{streb} have the usual physical meaning.

{\bf Remark} The negative spacetime divergence of the Distribution, obtained by pole taming, $\Tf=\Tf^{td} + \Tf^{rtd} + \Tf^{rd}$ 
is the well-known self-force, as it is shown in \cite{Mato}. The formulae there show, however, that 
both $\Tf^{td}$ and $\Tf^{rd}$ have zero divergence. Then it is seen clearly that the self-force is due to the interaction
between the tied field and the radiated field.

\subsection{Integral of point charges}

The conjecture in the previous subsection seems satisfactory for a point charge; but what about the radiated energy-momentum 
tensor of an arbitrary absolute current? This question, connected with the problem how to define a center of mass, 
emerged in \cite{Gral-Har-Wald} (p.9).

A particular answer is given as follows.

The absolute current of a point charge $e$ having the world line function $r$ is
$e\dot r\la_{\ran(r)}$, where $\la_{\ran r}$ is the Lebesgue measure of the world line $\ran r$ (given 
by the usual integration along the world line).

The absolute current consisting of point charges $e_1,\dots,e_n$ with world line functions $r_1,\dots,r_n$ is
$$\jj:=\sum_{a=1}^ne_a\dot r_a\la_{\ran r_a}.$$
Then the radiated field  is the sum of the radiated fields of the point charges,
$$ \F^{rd}:=\sum_a \F^{rd}_a$$
and the radiated energy-momentum tensor is
\begin{align*} \T^{rd}:=&-\F^{rd}\cdot\F^{rd} -\frac1{4}\text{Tr}\left(\F^{rd}\cdot\F^{rd}\right)\1= \\
=& \sum_a\sum_b\left( -\F^{rd}_a\cdot\F^{rd}_b -\frac1{4}\text{Tr}(\F^{rd}_a\cdot\F^{rd}_b)\1\right),\end{align*}

We conceive that any current consists of point charges flowing in spacetime, as it occurs
in our everyday electricity. This can be described by a simple generalization of the above formulae in 
such a way that there are given a set $A$, a measure $\alpha$ on $A$ and a point charge $e_a$ with world line 
function $r_a$ for each element $a$ of $A$; then
\beq\jj:=\intl_Ae_a\dot r_a \la_{\ran r_a}\ d\alpha(a)\enq
which is defined in such a way that for an arbitrary test function $\phi$
\beq (\jj\mid\phi):=\intl_A \left(\intl_\I(e_a\dot r_a(\s)\phi(r(\s))\ d\s\right) d\alpha(a).\enq

Then the radiated field $\F^{rd}$ of the current is the integral of the radiated fields of the point charges, 
defined by
\beq (\F^{rd}\mid\phi):=\intl_A (\F^{rd}_a\mid\phi) \ d\alpha(a),\enq
provided, of course, that the integral exists. The radiated energy-momentum tensor of the current is
\beq (\T^{rd}\mid\phi):=\intl_A\intl_A 
\left(-\F^{rd}_a\cdot\F^{rd}_b - \frac1{4}(\F^{rd}_a\cdot\F^{rd}_b)\1 \Bigm|\phi\right) \ d\alpha(a)\ d\alpha(b),\enq

It is a question, however, whether it suffices to consider absolute currents of this type only. The following 
answer could be encouraging.

A continuous absolute current is a function $x\mapsto\rho(x)\U(x)$ defined in spacetime, where 
$\rho$ is the absolute charge density and $\U$ is an absolute velocity field. 

\begin{all} The current $\rho\U$ is the integral of point charges if $\rho$ is constant along every world 
line function $r$ for which $\dot r(\s)=\U(r(\s))$ holds. \end{all}

{\bf Proof} We take a standard inertial frame $\uu$ with origin; the split form 
of the current is the function $(\ts,\q)\mapsto\Bigl(\rho_\uu(\ts,\q),\ \rho_\uu(\ts,\q)\vv_\uu(\ts,\q)\bigr)$
defined in the vectorized time and space of the frame where $\rho_\uu$ and $\vv_\uu$ are the charge density and 
the relative velocity according to $\uu$. 
For the sake of simplicity, the subscript $\uu$ will be omitted in what follows.

The Distribution corresponding to this split current is given by
\beq\label{huha}\intl_\I\intl_\Sd \rho(\ts,\q)\phi(\ts,\q)\ d\q\ d\ts, \qquad  
\intl_\I\intl_\Sd \rho(\ts,\q)\vv(\ts,\q)\phi(\ts,\q)\ d\q\ d\ts.\enq
                                                                                                                
Let $\ts\mapsto\rs_\ad(\ts)$ be the motion with respect to the frame for which $\rs_\ad'(\ts)=\vv(\ts,\rs_\ad(\ts))$  
and $\rs_\ad(0)=\ad\in\Sd$ hold.

The condition that $\rho$ is constant on the integral curves of $\vv$ is given by $\rho(\ts,\rs_\ad(\ts))=\rho(0,\ad)$ for all $\ts$.
The charge conservation involves  
$$\rho(\ts,\q)\ d\q = \rho(0,\ad)\ d\ad.$$ 

Then for every fixed $\ts$ taking the function $\ad\mapsto\rs_\ad(\ts)$ and the substitution  $\q=\rs_\ad(\ts)$ we have 
$$\rho(\ts,\rs_\ad(\ts))|\det\nabla_{(\ad)}\rs_\ad(\ts)|\ d\ad=\rho(0,\ad)\ d\ad$$
and the integrals in \eqref{huha} with the notation $e_\ad:=\rho(0,\ad)$ become
$$\intl_\I\intl_\Sd \rho(\ts,\rs_\ad(\ts))\phi(\ts,\rs_\ad(\ts))\ |\det\nabla_\ad\rs_\ad(\ts)|\ d\ad\ d\ts=
\intl_\I\intl_\Sd e_\ad\phi(\ts,\rs_\ad(\ts))\ d\ad\ d\ts,$$
\begin{multline*}
\intl_\I\intl_\Sd \rho(\ts,\rs_\ad(\ts))\vv(\ts,\rs_\ad(\ts))\phi(\ts,\rs_\ad(\ts))|\det\nabla_{(\ad)}\rs_\ad(\ts)|\ d\ad\ d\ts =\\
\intl_\I\intl_\Sd e_\ad\rs_\ad'(\ts)\phi(\ts,\rs_\ad(\ts))\ d\ad\ d\ts.\end{multline*}

Changing the order of integrations, we get the desired result: the continuous current is the integral of point charges.$\mqed$

It is worth noting that the requirement of the above proposition is equivalent to the velocity field being divergence-free.
Indeed, if $\rho$ is constant on the integral curves of $\U$, then $\U\cdot\Dd\rho=0$, and the charge conservation requires 
$0=\Dd\cdot\rho\U=\U\cdot\Dd\rho + \rho\Dd\cdot\U$. 

\section{Conclusions}

Energies and forces in electromagnetism have been examined by a systematic use of Distribution Theory and the
notions `extraneous' and `self'. 

The generally accepted and acceptable method in electrostatics has been reviewed on the basis that formulae valid for a 
system of point charges and dipoles one establishes formulae for continuous charge and dipole distributions. 

It has been revealed how those continuous formulae 
are usually transformed into another, questionable form by differentiation and integration. 

It has been pointed out that those transformed formulae,
which are valid only in the continuous case even with some conditions, are usually applied for point charges
which results in artificial problems. Such a most absurd problem is the infinite electric energy of a resting point charge. 

Those artificial problems have been ruled out. In particular, it can be unmistakably stated that the electrostatic energy of a point 
charge is zero.

The fundamental problems of energy and momentum balance equations have been demonstrated. 

First of all, the force density might be \eqref{lorplus} instead of the usually accepted  \eqref{lorero}. 

If dipoles and magnetic moments are not present then there is a simple balance-like equation but \eqref{endens} and 
\eqref{enaram} are not, in general, the electromagnetic energy density and energy flow density. In other 
words, \eqref{enmom} is not a real energy-momentum tensor, in general. A possible definition has been proposed 
for the radiated elecromagnetic energy-momentum. 

Lastly, we emphasize that point charges and dipoles as well as continuous charge and dipole distributions are models.
Considering a material object point-like, we do not assert that it is a point indeed; similarly, considering a 
ball with a continuous charge distribution, we do not assert that a real charge continuity exists there
because we know that an amount of discrete electrons establishes the charge distribution.

From this point of view, it is worth examining whether the stable, continuously charged ball 
with Poincar\'e force is a right model instead of a point-like charge.

\section{Appendix}                                                        

\subsection{Measures}

The Borel sets in spacetime or in the space of an observer are the elements of the $\sigma$-algebra 
generated by the open subsets.

Elementary notions of (vector) measures and integration by them can be found in \cite{Dinculeanu}.

In this paper a (vector) measure is defined (at least) on the bounded Borel sets.

\subsection{Distributions in spacetime} 

Spacetime test functions are smooth functions in $\aM$ or in $\M$ with compact support. 

A (scalar, vector, tensor) Distribution is a (scalar, vector, tensor) valued continuous linear map defined on test function.
The action of a Distribution $\Kf$ on a test function $\phi$ is denoted by $(\Kf\mid\phi)$. 

A (vector) measure is considered a  Distribution by integrating test functions.

Originally a Distribution acts on test functions but in some cases the action of a Distribution $\Kf$
on a locally integrable function $f$ can be defined as follows. 

A sequence  $n\mapsto\omega_n$ of test functions is said to converge to $f$ in the Distribution sense if 
$$\lim_{n\to\infty}\int \omega_n(x)\phi(x)\ dx = \int f(x)\phi(x)\ dx$$
for all test functions $\psi$. 

If $f=\lim_n\omega_n$ in the Distribution sense then
$$(\Kf\mid f):= \lim_{n\to\infty} (\Kf\mid \omega_n)$$
if the limit exists.

This definition is consistent: if $f$ is a test function itself then the limit exists and equals the action 
of $\Kf$ on $f$.

Similar notions hold, according to the sense, for test functions and Distributions in the affine space $\aS$ or vector 
space $\Sd$ of an inertial observer.

In particular, for all $0<a<b$ there is a test function $\omega_{a,b}$ in $\Sd$ such that
\beq\label{omeg}\omega_{a,b}(\q)\begin{cases}& = 1 \quad \text{if} \quad |\q|<a, \\
                                   &= 0 \quad \text{if} \quad |\q|>b, \\
                                    &\text{is between $0$ and $1$ if}  \quad a \le|\q|\le b.\end{cases},\enq
and $1=\lim_{n\to\infty}\omega_{na,nb}$ in the Distribution sense. 

\subsection{Three-dimensional integration}\label{4pi3}

Let $S_1(\0)$ be the unit sphere around zero in the three dimensional Euclidean vector space $\Sd$.

The radial coordinates of $\q\in\Sd$ are 
\beq r(\q):=|\q|\in \I^+_0, \quad \nb(\q):=\frac{\q}{|\q|}\in S_1(\0).\enq

Then the radial parametrization of $\Sd$ is 

\beq\label{radpar}(r,\nb)\mapsto r\nb.\enq

We applied the usual ambiguity that $r$ and $\nb$ denote functions in some respect and variables 
in an other respect.

Further, the usual parametrization of the unit sphere, given by an orthormal basis 
$\ee_1,\ee_2,\ee_3$ and the angles $\vartheta,\vf$, is

\beq\nb(\vartheta,\vf)= \ee_1\sin\vartheta\cos\vf + \ee_2\sin\vartheta\sin\vf + \ee_3\cos\vartheta.\enq
Then the following integral formula
\beq\intl_{S_1(\0)} f(\nb)\ d\nb = \intl_0^\pi\intl_{-\pi}^{\pi} f(\nb(\vartheta,\vf))\sin\vartheta\ d\vartheta\ d\vf\enq
is well known from which the equalities 

\beq\label{n1} \intl_{S_1(\0)} \nb \ d\nb=0, \qquad 
\intl_{S_1(\0)} \nb\otimes\nb\otimes\nb \ d\nb=0,\enq
\beq\label{nn} \intl_{S_1(\0)} \nb\otimes\nb \ d\nb=\frac{4\pi}{3}\1\enq
can be deduced where $\1$ is the identity of $\Sd$; and so on, the integral of even tensor products is a non-zero tensor,
the integral of odd tensor products is zero.

The other well known integral formula
\beq\label{radint1}\intl_\Sd f(\q)\ d\q =\intl_0^\infty r^2\intl_{S_1(\0)}f(r\nb)\ d\nb\ dr\enq
will be often used.

\subsection{Pole taming}\label{polet}

There is a well-defined method to attach a Distribution to a locally non-integrable function, having a pole
singularity in a Euclidean space. Originally the method was called regularization (\cite{Gelf-Shil}).
Since the Distribution obtained in this way is not regular, I call the method pole taming\footnote{\'Aron Szab\'o 
proposed this name}.

Only a special case is necessary for us, treated in a way slightly different from the original one. 

For the function $r(\q):=|\q|$ \ $(\q\in\Sd)$ in the three dimensional Euclidean vector space $\Sd$,
$\frac1{r^{2+m}}$ is not locally integrable if $m>0$. If $m$ is an even positive integer, then a Distribution 
can be attached to it by pole taming defined as follows.

For a test function $\psi$ let $T_\psi^{(m-1)}$ denote the Taylor polynomial of order $m-1$ at zero.
In a neighbourhood of zero, $\psi-T_\psi^{(m-1)}$ is a continuous function of order $r^m$, therefore  
$\frac{\psi-T_\psi^{(m-1)}}{r^{2+m}}$ is integrable there. Outside the support of $\psi$, however, the term 
containing the $(m-1)$-th derivative is not integrable.
 
$m$ is even, therefore the $(m-1)$-th derivative of $\psi$ at zero is an $(m-1)$-linear function of $\q=r\nb$,
in other words, it is a linear function of the $(m-1)$-th tensor power of $\q=r\nb$; thus,
integrating in the order given by \eqref{radint1}, this term will drop out and an integrable function remains.

We repeat for avoiding misunderstanding: it is known that, for an integrable function, the order of integration by 
$\nb$ and $r$ can be interchanged in \eqref{radint1}. If a function is not integrable, it can occur that the integral 
exists in one of the orders (but in the other order does not). Here we take advantage of the fact that the integral 
exists in the given order.
In this way the pole taming of $\frac1{r^{2+m}}$ results in the Distribution defined by 
\beq\label{tomp} \left(\tm\frac1{r^{2+m}}\Bigm|\psi\right):=
\intl_0^\infty\left(r^2\intl_{S_1(\0)}\frac{\psi(r\nb) - T_\psi^{(m-1)}(r\nb)}{r^{2+m}}\ d\nb\right)\ dr\ !\enq
where the exclamation mark calls attention to that the order of integration cannot be interchanged.

Later, for the sake of brevity, we also use the formula
\beq\label{tompi} \left(\tm\frac1{r^{2+m}}\Bigm|\psi\right):=
\intl_{\Sd}\frac{\psi(\q) - T_\psi^{(m-1)}(\q)}{|\q|^{2+m}}\ d\q\ !!\enq
where the double exclamation mark calls attention to that the integral must be taken in radial parametrization and in the 
given order.

We have to extend the notion of pole taming as follows. For a $k$-th tensor power $\nb^{\otm k}:=\nb\otm\otm\nb\otm\dots\nb$ 
we put
\beq\label{npol}\left(\tm\frac{\nb^{\otm k}}{r^{2+m}}\Bigm|\psi\right):=
\intl_0^\infty r^2\left(\intl_{S_1(\0)}\frac{\nb^{\otm k}
\bigl(\psi(r\nb) - T_\psi^{(m-1)}(r\nb\bigr)}{r^{2+m}}\ d\nb\right)\ dr\ !\enq
which makes sense if $m-1+k$ is odd; therefore 

-- \ if $m$ is even, then $k$ must be even, too,

-- \ if $m$ is odd, then $k$ must be odd, too.

More generally, if $\A(\nb)$ is a linear function of the tensor powers of $\nb$, such that
$$\intl_{S_1(\0)}\A(\nb)\otm\left(\nb^{\otm(m-1)}\right)\ d\nb=0,$$
then the definition \eqref{npol} is accepted with $\nb^{\otm k}$ replaced by $\A(\nb)$;
this can be meaningful for an odd or even $m$, depending on $\A$.

\section*{Acknowledgement} I am grateful to A. L\'aszl\'o for useful comments and to and P. V\'an for useful 
comments and for helping me to compose this article.

\end{document}